\documentclass[aip,author-numerical,jmp,showpacs]{revtex4-1}
\pdfoutput=1
\usepackage[sumlimits]{amsmath}
\usepackage{latexsym}
\usepackage{amssymb}
\usepackage{euscript}
\usepackage{graphicx}

\begin{document}
\tolerance=5000
\def\be{\begin{equation}}
\def\ee{\end{equation}}
\def\bea{\begin{eqnarray}}
\def\eea{\end{eqnarray}}
\def\bs{\boldsymbol}
\newcommand{\dslash}{d\!\!\!{}^{-}}
\renewcommand{\bs}{\boldsymbol}

\title{Topological discretization of bosonic strings}

\author{Gustavo Arciniega}
\email{gustavo.arciniega, fnettel, leopj@ciencias.unam.mx}
\author{Francisco Nettel}
\email{gustavo.arciniega, fnettel, leopj@ciencias.unam.mx} 
\author{Leonardo Pati\~no}
\email{gustavo.arciniega, fnettel, leopj@ciencias.unam.mx}
\affiliation{Departamento de F\'\i sica, Facultad de Ciencias, Universidad Nacional Aut\'onoma de M\'exico, \\
A.P. 50-542, M\'exico D.F. 04510, M\'exico}
\author{Hernando Quevedo}
\email{quevedo@nucleares.unam.mx} 
\affiliation{Instituto de Ciencias Nucleares, Universidad Nacional Aut\'onoma de M\'exico, \\
A.P. 70-543, M\'exico D.F. 04510, M\'exico}

\date{\today}

\begin{abstract}
We apply the method of topological quantization to obtain the bosonic string topological spectrum propagating on a flat background. We define the classical configuration of the system, and construct the corresponding principal fiber bundle (pfb) that uniquely represents it. The topological spectrum is defined through the characteristic class of the pfb. We find explicit expressions for the topological spectrum for particular configurations of the bosonic strings on a Minkowski background and show that they lead to a discretization of the total energy of the system.  
\end{abstract}

\maketitle
\section{Introduction}
\label{sec:int}

The main motivation to develop the method of topological quantization is to find an alternative to the ideas that prevail about the quantization of the gravitational fields. Nevertheless, as the method evolved we found ourselves exploring other classical theories with well established quantum counterparts, such as nonrelativistic quantum mechanics (finite number of degrees of freedom) and the bosonic string theory. So far, the canonical quantization, the most successful method to describe the discrete features of nature, still has to work out some answers for the theory of General Relativity. There are some unsolved challenges in the quantization of gravity, among them we find the problem of time, the lack of understanting of the ultimate meaning of the quantization of spacetime, the causality issues due to a fluctuant metric, the reconstruction problem and the appearance of nonrenormalizable divergences \cite{carlip, isham2}. At the present time, the main candidates for a quantum theory of gravity, i.e., string theory and loop quantum gravity, both use canonical quantization as it stands. We propose, with the method of topological quantization, to extract the discrete nature of physical systems without making assumptions or putting by hand any rule external to the geometric/topological structure that we use to represent the system under study.  

Dirac's idea  \cite{dirac1931} about the discretization of the relation between the charge of a magnetic monopole and a moving electron in the field generated by the former was the starting point to propose topological quantization as an alternative way of understanding the discrete nature of physical systems. In particular, the method was utilized to analyze the case of gravitational fields \cite{patquev2, patquev} and developed further for mechanical systems \cite{netque2, netquemo, netque}. The concept of topological quantization and the fundamental idea beneath these calculations has been broadly used in different contexts related to charge quantization in Yang-Mills theories \cite{deguchi}, instantons and monopoles configurations \cite{schwarz, zhong}, topological models of electromagnetism \cite{ranada}, current quantization of nanostructures \cite{bulgadaev} and in the theory of superconductors \cite{choi, leone}. 
Its relation to the cohomology theory has been also analyzed  \cite{alvarez}. Examples of topological quantization can also be 
found  in text books where its geometric formulation is applied to physical systems described by a hermitian line bundle \cite{frankel}. 
We will generalize this approach to include the case of an arbitrary physical system in the sense that we will provide a strict mathematical definition of classical configurations. Furthermore, the complete picture of topological quantization should also include the definition of states and its dynamical evolution in terms of geometric/topological structures, which, currently, is under research. In the beginning of this program we already established the geometric representation of the physical systems and from it we defined the topological spectrum.  

In the next section we briefly review some general aspects of the bosonic string theory. In section \ref{sec:tq} we give some elements of topological quantization and state the existence and unicity of the principal fiber bundle (pfb) that represents the physical system, followed by the general definition of topological spectrum. Section \ref{sec:gen} addresses the construction of the particular pfb and the definition of the topological spectrum for a bosonic string in a general background spacetime. In section \ref{sec:flat}, we turn our attention to the case of the bosonic string on a Minkowski background and its pfb. The analysis of the topological spectrum for some particular configurations is carried out.  Finally, in section \ref{sec:disc} we discuss our results and consider their implications over the embedding energy of the string.

\section{General aspects of the bosonic string}
\label{sec:bs}

The action integral for the free bosonic string moving in a general spacetime is given by the Nambu-Goto (N-G) action \cite{pol, joh}, which is proportional to the area of the worldsheet that describes the propagation of the string over a fixed background. To review this, consider a two-dimensional manifold $\mathcal{M}$ parametrized by $x^a$, $a=1,2$, and a $D$-dimensional manifold $\mathcal{N}$ with coordinates $X^\mu$, $\mu=0,\ldots, D-1$ and a metric tensor $\bs{G}$. Let $X:\mathcal{M} \to \mathcal{N}$ be a smooth map from $\mathcal{M}$, the worldsheet, to the spacetime $\mathcal{N}$. The induced metric on the embedded worldsheet is given by the pullback of $\bs{G}$ through the $X$ mapping, $\bs{g} = X^*\bs{G}$, whose components are,
\begin{equation}  \label{indmet}
g_{ab} = \frac{\partial X^\mu}{\partial x^a} \frac{\partial X^\nu}{\partial x^b} G_{\mu\nu}.
\end{equation}

Then, the N-G action is written out in terms of the induced metric as,
\begin{equation} \label{NG}
S_{NG} = -T\int d^2x \sqrt{|g|},
\end{equation}

\noindent where $T$ is the tension of the string and $g \equiv \det(g_{ab})$. The N-G action has two symmetries, the invariance under diffeomorphism on the worldsheet  $x'^a = x'^a(x)$ and the invariance under diffeomorphisms on the spacetime $X'^\mu = X'^\mu(X)$.

It is usual to start from an action, classically equivalent to (\ref{NG}), in which an auxiliary metric field $\bs{\gamma}$ is introduced on the worldsheet, 
\begin{equation}  \label{Polyakov}
S_P = -\frac{1}{4\pi\alpha'} \int d^2x \sqrt{|\gamma|} \,\gamma^{ab} g_{ab},
\end{equation}

\noindent where $\alpha'$ is related to the string tension by $T= \frac{1}{2\pi\alpha'}$. This is known as the Polyakov action \cite{pol, joh} and from a mathematical point of view is a harmonic map (or nonlinear sigma model) \cite{misner}. If we vary the Polyakov action with respect to the field $\bs{\gamma}$ we obtain the two-dimensional energy-momentum tensor $T_{ab} = \frac{4\pi}{\sqrt{\gamma}}\frac{\delta S_{P}}{\delta \gamma^{ab}}$ for the worldsheet, 
\begin{equation}   \label{gammag}
T_{ab} = g_{ab} - \frac{1}{2}\gamma^{cd}g_{cd}\gamma_{ab} = 0,
\end{equation}

\noindent which can be understood as a set of constraints that, among other things, suffice to prove the equivalence of (\ref{NG}) and (\ref{Polyakov}).

Varying with respect to $X^\mu$ the equations of motion that determine the dynamics of the string propagating in the spacetime follow,
\begin{equation}   \label{eqmot}
\frac{1}{\sqrt{|\gamma|}}\partial_a \left(\sqrt{|\gamma|}\, \gamma^{ab}\partial_b X^\mu \right) + \Gamma^{\mu}_{\;\alpha\beta}\, \gamma^{ab} \partial_a X^\alpha \partial_b X^\beta = 0,
\end{equation}

\noindent with $\partial_a \equiv \frac{\partial}{\partial x^a}$. When the background metric is $G_{\mu\nu} = \eta_{\mu\nu}$ the equations become,
\begin{equation}   \label{eqmotflat}
\partial_a \left(\sqrt{-\gamma}\,\gamma^{ab}\partial_b X^\mu \right) = 0.
\end{equation}

We are interested in exact solutions to  (\ref{eqmotflat}) as they will be necessary to find the induced metric which has a fundamental role in the explicit calculation of the topological spectrum. 

The Polyakov action possesses, besides the two symmetries of the N-G action, a third invariance under the Weyl transformation, a local rescaling of the metric tensor $\bs{\gamma}' = e^{\omega(x)}\boldsymbol{\gamma}$. In section \ref{sec:flat} we will analyze the general solution of (\ref{eqmotflat})  in order to find the topological spectrum for some configurations.

\section{Fundamentals of topological quantization}
\label{sec:tq}

A complete description of a physical system must include observables, states and its dynamical evolution. We further know that sometimes the observables have a discrete behavior. It is the aim of topological quantization to provide these three elements for any physical system from a geometric/topological outset and to find out if there is a discrete pattern in such description. Nowadays, we have established the first part of the method, which refers to the definition of the topological spectrum for some observables, meanwhile the definition of states and their dynamics remains as work in progress.  

We present here some basic elements for the definition of the topological spectrum. We define the classical configuration as a unique pair $(\mathcal{M}, \omega)$ composed by a Riemannian manifold $\mathcal{M}$ and a connection $\omega$ that represents the physical system. Uniqueness, in this case, means that two isomorphic manifolds with the same connection are identical classical configurations. As an example consider a gauge theory over a Minkowski spacetime $M_\eta$; this Riemannian manifold together with the connection one-form $A$, which takes values in the Lie algebra of a gauge group $G$, form the classical configuration. 

Furthermore, with the classical configuration we can build the pfb $\mathcal{P}$, using the Riemannian manifold $\mathcal{M}$  as the base space and the symmetry group of the theory $G$ as the structure group identical to the standard fiber. 

Given a local section $s_i$ which bears a local trivialization $(U_i, \phi_i)$ where $U_i \subset \mathcal{M}$ and $\phi_i : U_i \times G \to \pi^{-1}(U_i)$ \cite{nak}, it is possible to introduce a connection $\tilde{\omega}$ on $\mathcal{P}$ through the pullback $s_i^*\tilde{\omega} = \omega_i$ where $\omega_i$ is the connection $\omega$ on the open set $U_i$ in the base space $\mathcal{M}$. It can be shown that using these elements and the reconstruction theorem \cite{naber, nak} a unique principal fiber bundle exists which represents the physical system for the considered classical configuration. This has been done in the context of gravitational fields \cite{patquev} and for mechanical systems \cite{netquemo}. We shall show a similar result for the case in turn.

Once we have constructed the principal fiber bundle $\mathcal{P}$ from the classical configuration $(\mathcal{M}, \omega)$ the topological invariant properties of $\mathcal{P}$ can be used to characterize the physical system. This can be done employing the characteristic class of the pfb $C(\mathcal{P})$, that integrated over a cycle of $\mathcal{M}$ constitutes also an invariant of the bundle. The characteristic class $C(\mathcal{P})$, properly normalized leads to,
\begin{equation}
\int C(\mathcal{P}) = n,
\end{equation}

\noindent where $n$ is an integer called the characteristic number \cite{damas}. For the cases we analyze, the symmetry group of the theory may be reduced to an orthogonal group $SO(k)$ by introducing an orthonormal frame on the base manifold. Then, the characteristic class for such bundles is the Pontrjagin class $p(\mathcal{P})$, or the Euler class $e(\mathcal{P})$ in case of $k$ being an even integer. These characteristic classes can be spelled out in terms of the curvature two-form $R$ of the base space by means of the polynomials invariant under the action of the structure group $SO(k)$ \cite{nashsen},
\begin{equation}
\det \left(It - \frac{R}{2\pi}\right) = \sum_{j=0}^{k} p_{k-j}(R)t^j.
\end{equation}

The Euler class $e(\mathcal{P})$, only defined for even $k$, is expressed in terms of the curvature two-form $R$ of a (pseudo-)Riemannian connection on the base space as \cite{damas, nashsen},
\begin{equation}  \label{eulergen}
e(\mathcal{P}) = \frac{(-1)^m}{2^{2m}\pi^m m!}\epsilon_{i_1 i_2 \cdots i_{2m}} R^{i_1}_{\;i_2} \wedge R^{i_3}_{\;i_4} \wedge \cdots \wedge R^{i_{2m-1}}_{\;i_{2m}},
\end{equation}

\noindent where $2m=k$. It is clear that being in terms of the curvature form, the characteristic classes depend on some parameters $\lambda_i$, $i=1,\ldots,s$ which bear physical information of the system; thus, once we integrate the characteristic class, we end up with a discrete relation for $\lambda_i$,
\begin{equation}
\int C(\mathcal{P}) = f(\lambda_1, \ldots, \lambda_s) = n,
\end{equation}

\noindent where $n \in \mathbb{Z}$. This relationship is what we define as the topological spectrum and constitutes a discretization for some of the parameters determining the properties of the physical system of interest. In the next section we will explore in detail these definitions and the existence and uniqueness of the pfb for the bosonic string system.

\section{Bosonic string on a general background}
\label{sec:gen}

In this section we construct explicitly the principal fiber bundle for the bosonic string on a general background. It is natural in this case to consider the worldsheet $\mathcal{M}$ embedded in the spacetime $\mathcal{N}$ as the base space provided with the induced metric $\bs{g} = X^*\bs{G}$. Hence, the classical configuration is $(\mathcal{M}_{\bs{g}}, \omega)$, where $\omega$ is the Levi-Civita connection on $\mathcal{M}$ compatible with $\bs{g}$. We take the invariance under diffeomorphisms on the worldsheet as the structure group (isomorphic to the standard fiber), since this is the fundamental symmetry of the two dimensional action integral. 

The group of diffeomorphisms on $\mathcal{M}$ can be reduced to the orthogonal group by introducing a semiorthonormal frame. Indeed, given $\{e_i\}$ with $i = 1, 2$, an orientable orthonormal frame on $\mathcal{M}$ such that $\bs{g}(e_i, e_j) = \eta_{ij}$, two distinct bases are related by an orthogonal transformation, $e'_i = e_j(\Lambda^{-1})^{j}_{\;\;i}$, where $\Lambda \in SO(1,1)$. There is a one-form basis $\{\theta^i\}$ dual to the orthonormal frame from which it is possible to express the induced metric tensor as $\bs{g} = \eta_{ij}\; \theta^i \!\otimes \theta^j$; thus, the reduction of the symmetry group to $SO(1,1)$ is accomplished. Therefore, the principal fiber bundle $\mathcal{P}$ can be constructed from the classical configuration $(\mathcal{M}_{\bs{g}},\omega')$, with $\omega'$ the spin connection taking values in the Lie algebra $so(1,1)$, and $SO(1,1)$ as the structure group. This is summarized in the following result:

\textbf{Theorem:} A bosonic string propagating in a general background $(\mathcal{N}, \bs{G})$ described by the Nambu-Goto action can be represented by a unique principal fiber bundle $\mathcal{P}$, with the semi-Riemannian manifold $(\mathcal{M}, \bs{g})$ as the base space, $SO(1,1)$ as the structure group (identical to the standard fiber) and with a $\bs{g}-$compatible connection $\bs{\omega}$ which takes values in the Lie algebra $so(1,1)$.

The proof of this theorem is completely analogous to the one that appears in previous works \cite{patquev, netquemo} and we refer the reader to them for the details. It should be sufficient to mention that it rests on the reconstruction theorem for fiber bundles \cite{naber}.

The Euler characteristic class for the principal fiber bundle $\mathcal{P}$ with a two-dimensional base space and $SO(1,1)$ as the structure group reduces from (\ref{eulergen}) to
\begin{equation}  \label{Eulerform}
e(\mathcal{P}) =  -\frac{1}{2\pi} R^{1}_{\;2}.
\end{equation}
In the conformal gauge, using coordinates $\{\tau,\sigma\}$ in Eq.(\ref{indmet}), the worldsheet metric 
 turns out to be conformal to the two-dimensional Minkowski metric, $\bs{g} = g_{\sigma\sigma}\bs{\eta}$. 
 
In this gauge the Euler characteristic class takes the following explicit form,
\begin{equation}   \label{Eulerformalt}
e(\mathcal{P}) = -\frac{1}{4\pi} \left[ \partial_\tau \left(\frac{1}{g_{\sigma\sigma}}\partial_\tau g_{\sigma\sigma} \right) - \partial_\sigma \left(\frac{1}{g_{\sigma\sigma}}\partial_\sigma g_{\sigma\sigma} \right) \right] d\tau \wedge d\sigma.
\end{equation}

Consequently, the determination of the topological spectrum reduces to the computation of the conformal factor $g_{\sigma\sigma}$ 
and the integral $\int C(\mathcal{P}) = n \in \mathbb{Z}$, {\it regardless of the background metric}. This shows for this particular case that the formalism of topological quantization is background independent.  We will use this property in the following sections
to determine specific topological spectra on diverse backgrounds.

\section{Bosonic string on a Minkowski background}
\label{sec:flat}

The worldsheet that minimizes the action of a bosonic string propagating in a flat background, $G_{\mu\nu} = \eta_{\mu\nu}$ is described by the set of embedding functions $\{X^\mu \}$ satisfying the equations of motion 
\begin{equation}  \label{wave}
\left(-\partial_{\tau}^{2} + \partial_{\sigma}^{2}\right)X^\mu(\tau,\sigma) = 0,
\end{equation}

\noindent with general solution
\begin{equation}  \label{fyg}
X^\mu(\tau, \sigma) = F^\mu(\tau + \sigma) + G^\mu(\tau - \sigma).
\end{equation}

We have chosen the conformal gauge to write these and the forthcoming expressions. The set (\ref{gammag}) of constraint equations takes the form
\begin{align}  \label{const1}
\left( \partial_\tau X^\mu \partial_\tau X^\nu + \partial_\sigma X^\mu \partial_\sigma X^\nu \right)\eta_{\mu\nu} &= 0,  \nonumber \\
\partial_\tau X^\mu \partial_\sigma X^\nu \eta_{\mu\nu} &= 0,
\end{align}

\noindent from which the conformal factor of the induced metric can be computed. Let us see how in the case of a Minkowski background choosing the light cone gauge leaves no residual gauge freedom. Consider a whole class of gauges given by \cite{zwiebach},
\begin{align} \label{clasenormas}
\hat{n}\cdot X(\tau,\sigma) &= \beta \alpha' (\hat{n}\cdot p)\tau, \nonumber  \\
(\hat{n}\cdot p)\sigma &= \frac{2\pi}{\beta}\int_0^{\sigma} d\tilde{\sigma}\,\, \hat{n}\cdot \!P^\tau(\tau,\tilde{\sigma}), 
\end{align}

\noindent where $\hat{n}$ is a unitary vector which fixes the relation between the parameters of the worldsheet with the spacetime coordinates, and $\hat{n}\cdot X = \hat{n}^\mu X^\nu \eta_{\mu\nu}$. The constant $\beta$ determines whether we are dealing with an open ($\beta=2$) or closed ($\beta=1$) string; $P^\tau$ is the momentum density along the string, and $p$ the four momentum. Using light cone coordinates for the background space,
\begin{align}
X^+ &= \frac{X^0 + X^1}{\sqrt{2}}, \nonumber \\
X^- &= \frac{X^0 - X^1}{\sqrt{2}}, \nonumber \\
X^I &= X^I, \quad \text{con} \quad I = 2,\ldots, D-1,
\end{align}

\noindent the line element for the Minkowski spacetime takes the following form
\begin{equation} \label{dsGnulas}
ds_G^2 = -2dX^+dX^- + dX^IdX^J\delta_{IJ}.
\end{equation}

The light cone gauge is fixed choosing the unitary vector $\hat{n}$ as,
\begin{equation}
n^\mu = \left(-\frac{1}{\sqrt{2}},\frac{1}{\sqrt{2}},0,\ldots,0\right).
\end{equation}

Then, the equations (\ref{clasenormas}) that determine this specific gauge read,
\begin{align} \label{sconodeluz}
X^+(\tau,\sigma) &= \beta\alpha' p^+ \tau, \nonumber \\
p^+ \sigma &= \frac{2\pi}{\beta} \int_0^{\sigma} d\tilde{\sigma} P^\tau{}^+(\tau,\sigma),
\end{align}

\noindent where $\hat{n} \cdot P^\tau$ is constant along the string and consequently $p^+$ too, and we notice that the gauge is completely fixed.

From this we also see that the parameter $\sigma$ takes values in the interval $[0, 2\pi]$ for a closed string (periodic boundary conditions). From the constraints equations (\ref{const1}) in this gauge,
\begin{align}  \label{constconoluz}
\partial_\tau X^- &= \frac{1}{2\alpha'p^+}\left(\partial_\tau X^I\partial_\tau X^J + \partial_\sigma X^I\partial_\sigma X^J\right)\delta_{IJ}, \nonumber \\
\partial_\sigma X^- &= \frac{1}{\alpha'p^+}\partial_\tau X^I \partial_\sigma X^J \delta_{IJ},
\end{align}

\noindent we observe that the component $X^-$ can be found once the transverse sector $X^I(\tau,\sigma)$, $I= 2, \ldots, D-1$, is solved; therefore, it does not represent a dynamical degree of freedom.  

To obtain the topological spectrum integrating the Euler form (\ref{Eulerformalt}), we must first find the conformal factor of the induced metric $g_{\sigma\sigma}$, which in view of the constraints (\ref{constconoluz}) reduces to 
\begin{equation}  \label{gssconoluz}
g_{\sigma\sigma} = \partial_\sigma X^I \partial_\sigma X^J \delta_{IJ}.
\end{equation}

It is clear now that the conformal factor only depends on the dynamics of the string, that is, the transverse sector $X^I$ for the solution to the equations of motion.

\subsection{Topological spectrum for the closed bosonic string}
\label{sub:closedst} 
    
In this section we will obtain the topological spectrum for some particular configurations (solutions) of the closed bosonic string. In this case periodic boundary conditions must be imposed \cite{joh},
\begin{align}
X^\mu(\tau,\sigma_1) &= X^\mu(\tau, \sigma_2)  \\
\partial_\sigma X^\mu(\tau,\sigma_1) &= \partial_\sigma X^\mu(\tau,\sigma_2) \\\gamma_{ab}(\tau,\sigma_1) &= \gamma_{ab}(\tau,\sigma_2),
\end{align} 

\noindent where $\sigma_1 = 0$ and $\sigma_2 = 2\pi$. The solutions are described through two sets of oscillation modes, which are usually interpreted as left moving $\{\tilde{\alpha}^{\mu}_{k}\}$ and right moving $\{\alpha^{\mu}_{k}\}$ waves along the string \cite{pol}. In the conformal gauge the solutions may be expressed as,
\begin{multline}  \label{cssol}
X^\mu(\tau,\sigma) = x^{\mu}_{0} + \sqrt{2\alpha'} \alpha^{\mu}_{0}\tau + \sqrt{\frac{\alpha'}{2}}\sum_{k = 1}^{\infty} \frac{1}{\sqrt{\omega_k}} \left( \alpha^{\mu}_{k} e^{-i\omega_k(\tau-\sigma)}  \right. \\
\left. +  \alpha^{\mu}_{k}{}^* e^{i\omega_k(\tau-\sigma)} + \tilde{\alpha}^{\mu}_{k} e^{-i\omega_k(\tau+\sigma)} + \tilde{\alpha}^\mu_k{}^* e^{i\omega_k(\tau+\sigma)} \right),
\end{multline}

\noindent where, here and throughtout this section, $\mu = 0,\ldots, D-1$ and $\omega_k = k$. The periodicity in $\sigma$ has been considered, leading to the condition that the zero modes are equal, $\tilde{\alpha}^{\mu}_{0} = \alpha^{\mu}_{0}$. The constraints can also be expressed as two independent sets of equations in terms of the modes of oscillation
\begin{equation} \label{csconst}
L_k = \frac{1}{2} \sum_{p \in \mathbb{Z}} \alpha^{\mu}_{p - k} \alpha^{\nu}_{p} \eta_{\mu\nu} = 0 \qquad \tilde{L}_k = \frac{1}{2} \sum_{p \in \mathbb{Z}} \tilde{\alpha}^{\mu}_{p - k} \tilde{\alpha}^{\nu}_{p} \eta_{\mu\nu} = 0.
\end{equation}

In the light cone gauge the dynamical fields $X^I(\tau,\sigma)$ take the same form as above (\ref{cssol}), just considering the transverse index $I$ instead of the spatiotemporal $\mu$. Only these transverse fields enter in the expression for the conformal factor $g_{\sigma\sigma}$. We introduce the polar notation for the modes coefficients, $\alpha_k^I = r_k^I e^{-i\gamma_k^I}$ and $\tilde{\alpha}_k^I = \tilde{r}_k^I e^{-i\tilde{\gamma}_k^I}$, such that the solutions for the transverse fields are written as
\begin{multline}
X^I(\tau,\sigma) = x_0^I + \sqrt{2\alpha'}\alpha_0^I \tau  \\ 
+ \sqrt{2\alpha'} \sum_{k=1}^\infty \frac{1}{\sqrt{\omega_k}} \left[ r_k^I \cos \omega_k(\tau -\sigma + \gamma_k^I) + \tilde{r}_k^I \cos \omega_k(\tau +\sigma + \tilde{\gamma}_k^I) \right].
\end{multline}   

Then, the metric function $g_{\sigma\sigma}$ which determines the Euler characteristic class (\ref{Eulerformalt}) in this gauge is given in general by a infinite sum of oscillation modes,
\begin{multline}
g_{\sigma\sigma}(\tau,\sigma) = 2\alpha'\sum_{k,l=1}^{\infty} \sqrt{\omega_k \omega_l} \bigg[r_k^I \sin \omega_k(\tau -\sigma + \gamma_k^I) - \tilde{r}_k^I \sin \omega_k(\tau +\sigma + \tilde{\gamma}_k^I)\bigg]   \\
\times \bigg[r_l^J \sin \omega_l(\tau -\sigma + \gamma_l^J) - \tilde{r}_l^J \sin \omega_l(\tau +\sigma + \tilde{\gamma}_l^J)\bigg]\delta_{IJ}.
\end{multline}

It then follows that the integration of the corresponding topological invariant involves the manipulation of 
 infinite series with the consequent  technical difficulties.
Hence, we take into account particular configurations with only a few nonvanishing modes of oscillation that allow us to reach concrete expressions for their spectra.  

\subsection{Topological spectrum of particular configurations}
\label{sub:partconf}

To investigate how the interaction of different modes of oscillation affects the geometric properties of the underlying pfb, let us consider the case of a right mode $\alpha_k^{J_1} \neq 0$ in the direction $J_1$, and a left mode in a different direction $J_2$, $\tilde{\alpha}_l^{J_2} \neq 0$. The transverse fields that involve these modes of oscillation are
\begin{align}   \label{solcasomin}
X^{J_1} &= x_0^{J_1} + \sqrt{2\alpha'}\alpha_0^{J_1}\tau + \sqrt{2\alpha'}\,\frac{r_k^{J_1}}{\sqrt{\omega_k}} \cos \,\omega_k(\tau -\sigma + \gamma_k^{J_1}),  \nonumber \\
X^{J_2} &= x_0^{J_2} + \sqrt{2\alpha'}\alpha_0^{J_2}\tau + \sqrt{2\alpha'}\,\frac{\tilde{r}_l^{J_2}}{\sqrt{\omega_l}} \cos \,\omega_l(\tau +\sigma + \gamma_l^{J_2}),
\end{align}
\noindent where we have expressed the coefficients in the polar notation. In all the other transverse directions $J \neq J_1, J_2$, the fields describe only the motion of the center of mass, $X^J(\tau,\sigma) = x_0^J + \sqrt{2\alpha'} \alpha_0^J \tau$. The conformal factor for the induced metric is,
\begin{equation}
g_{\sigma\sigma} = 2\alpha' \left[\omega_k (r_k^{J_1})^2 \sin^2 \omega_k(\tau -\sigma +\gamma_k^{J_1}) + \omega_l (\tilde{r}_l^{J_2})^2 \sin^2 \omega_l(\tau +\sigma +\gamma_l^{J_2}) \right],
\end{equation}
\noindent and the Euler characteristic class,
\begin{equation}  \label{eulercasomin}
e(\mathcal{P}) = \frac{\omega_k^2 \omega_l^2 (r_k^{J_1} \tilde{r}_l^{J_2})^2 \sin 2\omega_k(\tau -\sigma +\gamma_k^{J_1}) \sin 2\omega_l(\tau +\sigma +\tilde{\gamma}_l^{J_2})}{\pi \left[ (r_k^{J_1})^2 \sin^2 \omega_k(\tau -\sigma +\gamma_k^{J_1}) + (\tilde{r}_2^{J_2})^2 \sin^2 \omega_l(\tau +\sigma +\tilde{\gamma}_l^{J_2}) \right]^2} d\tau \wedge  d\sigma.
\end{equation}

In order to integrate the Euler form (\ref{eulercasomin}) we must specify the limits in the domain of integration. For the parameter $\sigma$ the interval is $[0,2\pi]$ and is fixed, while for $\tau$ we notice that the above expression is periodic in this parameter and we may choose a complete cycle. To perform the integral it is convenient to use null-like coordinates patches that cover the entire domain of integration, for the details of the calculation we refer the reader to the appendix \ref{app:uno}. 
In this case, it turns out that the integral of the Euler class vanishes identically, meaning that no discrete relation between the parameters $r_k^{J_1}$ and $\tilde{r}_l^{J_2}$ is established. This is so due to the lack of interaction between the modes as they point in perpendicular directions of the background spacetime. 

Next we calculate the topological spectrum for the string with two nonvanishing modes of oscillation in the same transverse direction, that is, a right $k$-mode $\alpha_k^J$ and a left $l$-mode $\tilde{\alpha}_l^J$. The transverse field in the relevant direction $I=J$ is
\begin{multline} \label{soltwomodes}
\quad X^J(\tau,\sigma) = x_0^J + \sqrt{2\alpha'}\alpha_0^J \tau  \\ 
+ \sqrt{2\alpha'}\left[\frac{r_k}{\sqrt{\omega_k}} \cos \omega_k(\tau -\sigma +\gamma_k) + \frac{\tilde{r}_l}{\sqrt{\omega_l}} \cos \omega_l(\tau +\sigma + \tilde{\gamma}_l) \right],
\end{multline}

\noindent where we have used again the polar notation, $\alpha_k^J = r_k e^{-i\gamma_k}$ y $\tilde{\alpha}_l^J = \tilde{r}_l e^{-i\tilde{\gamma}_l}$. In all the remaining directions, $I \neq J$, the solutions describe the motion of the center of mass and only depend on $\tau$.

The conformal factor is given by
\begin{equation}
g_{\sigma\sigma}(\tau,\sigma) = 2\alpha \left[\sqrt{\omega_k}\, r_k \sin \omega_k(\tau -\sigma + \gamma_r) - \sqrt{\omega_l}\, \tilde{r}_l \sin \omega_l(\tau +\sigma + \tilde{\gamma}_l) \right]^2,
\end{equation}

\noindent and the Euler form as
\begin{equation}  \label{eulercasonomintau}
e(\tau,\sigma) = -\frac{2(\omega_k\omega_l)^{\frac{3}{2}}\,r_k\tilde{r}_l \, \cos \omega_k(\tau -\sigma +\gamma_k) \cos \omega_l(\tau +\sigma +\tilde{\gamma}_l)}{\pi \left[\sqrt{\omega_k}\, r_k \sin \omega_k(\tau -\sigma +\gamma_k) - \sqrt{\omega_l}\,\tilde{r}_l \sin \omega_l(\tau +\sigma +\tilde{\gamma}_l) \right]^2} \,\, d\tau \wedge d\sigma.
\end{equation}

To obtain the topological spectrum we must integrate this expression for $\sigma \in [0,2\pi]$ and a period in $\tau$. 

We use the coordinate transformation (\ref{coordcasomin}) and cover the region of integration as explained in Appendix \ref{app:uno}.
For the regions $I$ and $IV$ the Euler characteristic class takes the following form
\begin{equation}
e(x,y) = -\frac{1}{\pi} \frac{r_k \tilde{r}_l\,\sqrt{\omega_k\omega_l}}{\left(\sqrt{\omega_k}\,r_k x - \sqrt{\omega_l}\,\tilde{r}_l y \right)^2} \,\, dx \wedge dy,
\end{equation}

\noindent and for the type $II$ and $III$ we have
\begin{equation}
e(x,y) = -\frac{1}{\pi} \frac{r_k \tilde{r}_l\,\sqrt{\omega_k\omega_l}}{\left(\sqrt{\omega_k}\, r_k x + \sqrt{\omega_l}\, \tilde{r}_l y \right)^2} \,\, dx \wedge dy.
\end{equation}

\begin{figure}[htb]
\centering
\includegraphics[scale=0.15]{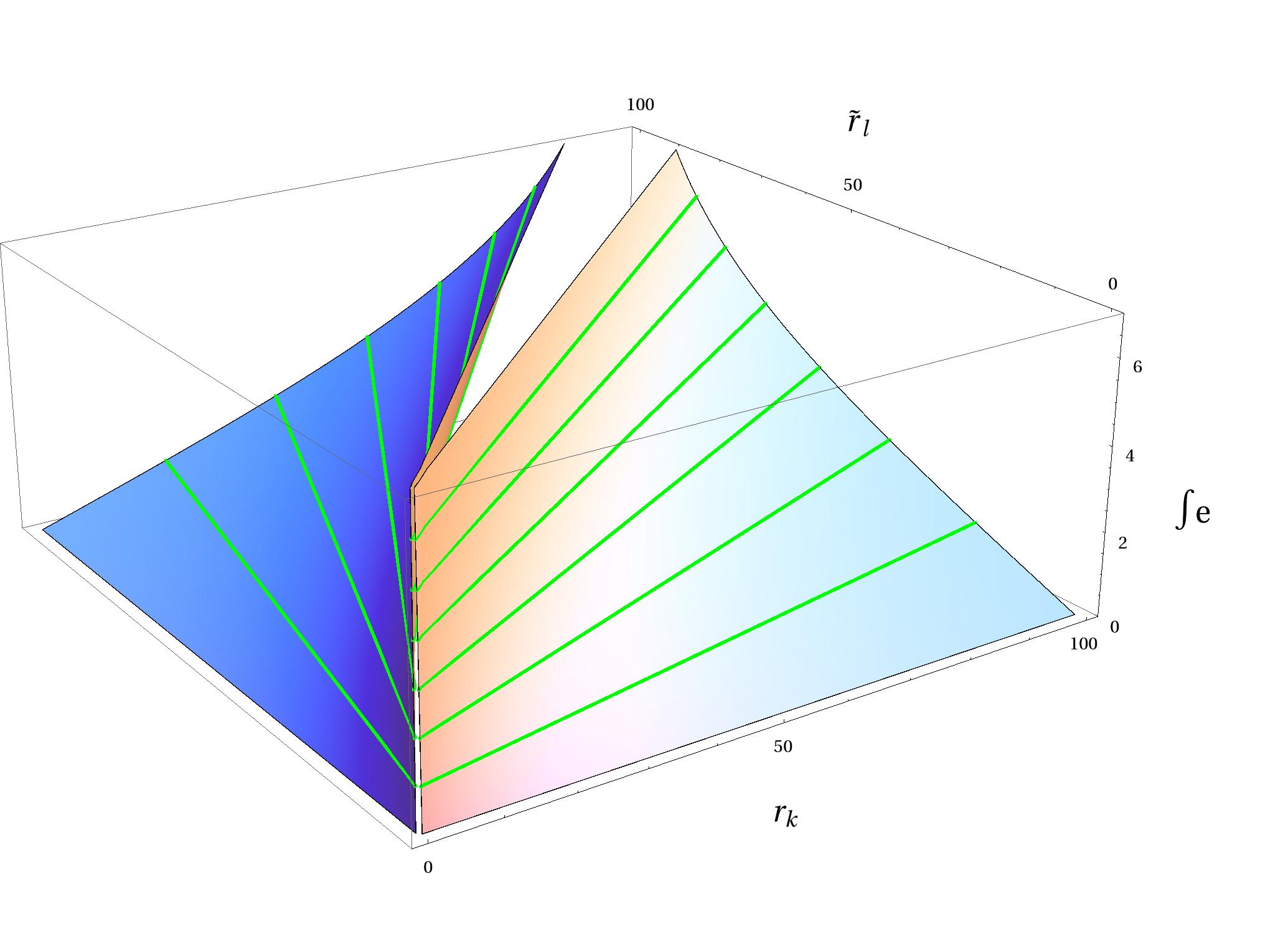}\caption{Illustration of the topological spectrum in the case of two nonvanishing modes of oscillation (right and left) that point in the same spacetime direction. The lines on the surface show the simultaneous values of $r_k$ and $\tilde{r}_l$ that are permitted by the discrete relation for $k=\omega_k=1$ and $l=\omega_l=1$.}
\label{casonominespe}
\end{figure}

The outcome of the integration yields a discrete relation between the amplitudes of the oscillation modes $r_k$ and $\tilde{r}_l$,
\begin{equation}   \label{casonominspec1}
\frac{4}{\pi}\, \omega_k\omega_l\, \ln \left[\frac{\left(\sqrt{\omega_l}\,\tilde{r}_l + \sqrt{\omega_k}\,r_k\right)^2}{\left(\sqrt{\omega_l}\,\tilde{r}_l - \sqrt{\omega_k}\, r_k\right)^2} \right] = n,
\end{equation}

\noindent where $n$ is an integer. This is the topological spectrum for the case of two nonvanishing modes of oscillation (right and left) in the same direction of the flat background space. We show in figure \ref{casonominespe} the allowed values for  $r_k$ and $\tilde{r}_l$ according to the relation (\ref{casonominspec1}).  

Now we can add another nonvanishing mode of oscillation to the ones we had in the previous case. Then, there are two modes in the direction $I=J_1$, $\alpha_k^{J_1}$, $\tilde{\alpha}_l^{J_1}$ right and left respectively, and a third right $k$-mode in a independent direction $I=J_2$, $\alpha_k^{J_2}$ (the case of including a left mode instead can be treated in a similar fashion). The relevant transverse fields are,
\begin{multline}
\quad X^{J_1}(\tau,\sigma) = x_0^{J_1} + \sqrt{2\alpha'}\alpha_0^{J_1}   \\ 
+ \sqrt{2\alpha'} \left[\frac{r_k^{J_1}}{\sqrt{\omega_k}} \cos \omega_k(\tau -\sigma + \gamma_k^{J_1}) + \frac{\tilde{r}_l^{J_1}}{\sqrt{\omega_l}} \cos \omega_l(\tau +\sigma +\tilde{\gamma}_l^{J_1}) \right],
\end{multline}
\begin{equation}
X^{J_2}(\tau,\sigma) = x_0^{J_2} + \sqrt{2\alpha'}\alpha_0^{J_2} + \sqrt{2\alpha'} \frac{r_k^{J_2}}{\sqrt{\omega_k}} \cos \omega_k(\tau -\sigma +\gamma_k^{J_2}).
\end{equation}

Integrating as the two former cases we obtain the topological spectrum which generalizes the relation (\ref{casonominspec1})
\begin{equation}\label{ts2}
\frac{4}{\pi}\,\omega_k\omega_l\, \ln \left[ \frac{\omega_k\left(r_k^{J_2}\right)^2 + \left(\sqrt{\omega_k}r_k^{J_1} + \sqrt{\omega_l}\tilde{r}_l^{J_1}\right)^2}{\omega_k\left(r_k^{J_2}\right)^2 + \left(\sqrt{\omega_k}r_k^{J_1} - \sqrt{\omega_l}\tilde{r}_l^{J_1}\right)^2} \right] = n.
\end{equation}

If a left $l$-mode in the $I=J_2$ direction is included to the modes of the preceding case we obtain the following discrete relation,   
\begin{equation}\label{ts3}
\frac{4}{\pi}\, \omega_k\omega_l\, \ln \left[ \frac{\left(\sqrt{\omega_k}r_k^{J_2} + \sqrt{\omega_l}\tilde{r}_l^{J_2}\right)^2 + \left(\sqrt{\omega_k}r_k^{J_1} + \sqrt{\omega_l}\tilde{r}_l^{J_1}\right)^2}{\left(\sqrt{\omega_k}r_k^{J_2} - \sqrt{\omega_l}\tilde{r}_l^{J_2}\right)^2 + \left(\sqrt{\omega_k}r_k^{J_1} - \sqrt{\omega_l}\tilde{r}_l^{J_1}\right)^2} \right] = n,
\end{equation}

\noindent giving the guideline to generalize the topological spectrum to the case in which there are two or more nonvanishing modes (right and left) in each spacetime direction.

\subsection{Discretization of the energy}

Let us now find out how the restrictions imposed by the topological spectrum reflect on a physical quantity such as the Hamiltonian function. We shall do this for the particular configuration described by the solutions (\ref{soltwomodes}) that lead to the relation (\ref{casonominspec1}). The Hamiltonian density in the light cone gauge is \cite{joh}
\begin{equation}
\mathcal{H} = \frac{1}{4\pi\alpha'} \left[ \partial_\tau X^K \partial_\tau X^L + \partial_\sigma X^K \partial_\sigma X^L  \right] \delta_{KL}, 
\end{equation}
so that the Hamiltonian function $H = \int_0^{2\pi} \mathcal{H} d\sigma$ for this particular configuration is
\begin{equation}  \label{hamil}
H = H_0 + \omega_k r_k^2 + \omega_l \tilde{r}_l^2, \quad H_0 = \sum_K \left(\alpha_0^K\right)^2.
\end{equation}

On the other hand, from the topological spectrum (\ref{casonominspec1}) we can derive an expression for the
term $\omega_k r_k^2 + \omega_l \tilde{r}_l^2$ which, when replaced in the above Hamiltonian, yields
\begin{equation}  \label{hamil1}
H = H_0 - 2 \sqrt{\omega_k\omega_l}\, r_k \tilde{r}_l \left( \frac { 1 + e^{n/\omega_{kl}}}{1 - e^{n/\omega_{kl}}  }\right),
\quad  \omega_{kl} = \frac{4}{\pi} \omega_k\omega_l \ ,
\end{equation}
or, equivalently for $\sqrt{\omega_k} r_k > \sqrt{\omega_l}\tilde r _l$,
\begin{equation}  \label{hamil2}
H = H_0 + \omega_k r_k^2 \left[1+\left(\frac{1-e^{n/2\omega_{kl}}}{1+e^{n/2\omega_{kl}}}\right)^2\right]
\ ,
\end{equation} 

\noindent and for $\sqrt{\omega_k} r_k < \sqrt{\omega_l}\tilde r _l$,
\begin{equation}  \label{hamil3}
H = H_0 + \omega_k r_k^2 \left[1+\left(\frac{1+e^{n/2\omega_{kl}}}{1-e^{n/2\omega_{kl}}}\right)^2\right]
\ .
\end{equation} 

We conclude that the topological quantization leads to a discrete Hamiltonian function. In fact, for any given bosonic string configuration, which corresponds to fixed values of the frequencies and amplitudes, the Hamiltonian can take only those values that are allowed by the discrete relationship (\ref{hamil2}, \ref{hamil3}) which depends explicitly on the integer $n$. This is the main result of our analysis. 

\begin{figure}[htb]
\centering
\includegraphics[scale=0.06]{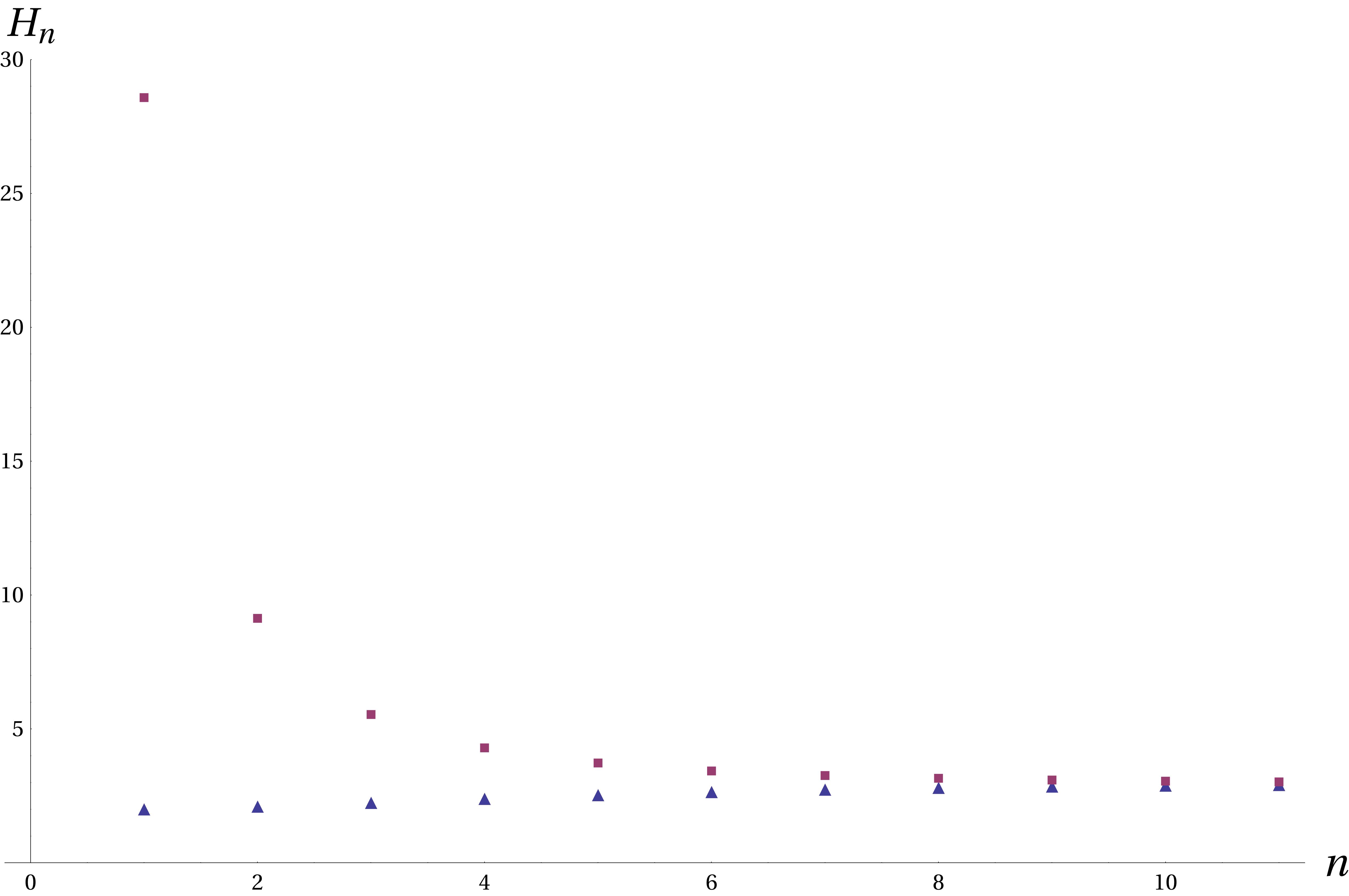}\caption{Graphic of the Hamiltonian function with $k=\omega_k=1$, $l=\omega_l=1$, $H_0=1$ and $r_k = 1$, showing a discrete behavior. The cases $\sqrt{\omega_l}\tilde{r}_l > \sqrt{\omega_k}r_k$ and $\sqrt{\omega_l}\tilde{r}_l < \sqrt{\omega_k}r_k$ are indicated by squares and triangles, respectively.}\label{tophamilfigmesh}
\end{figure}

\begin{figure}[htb]
\centering
\includegraphics[scale=0.20]{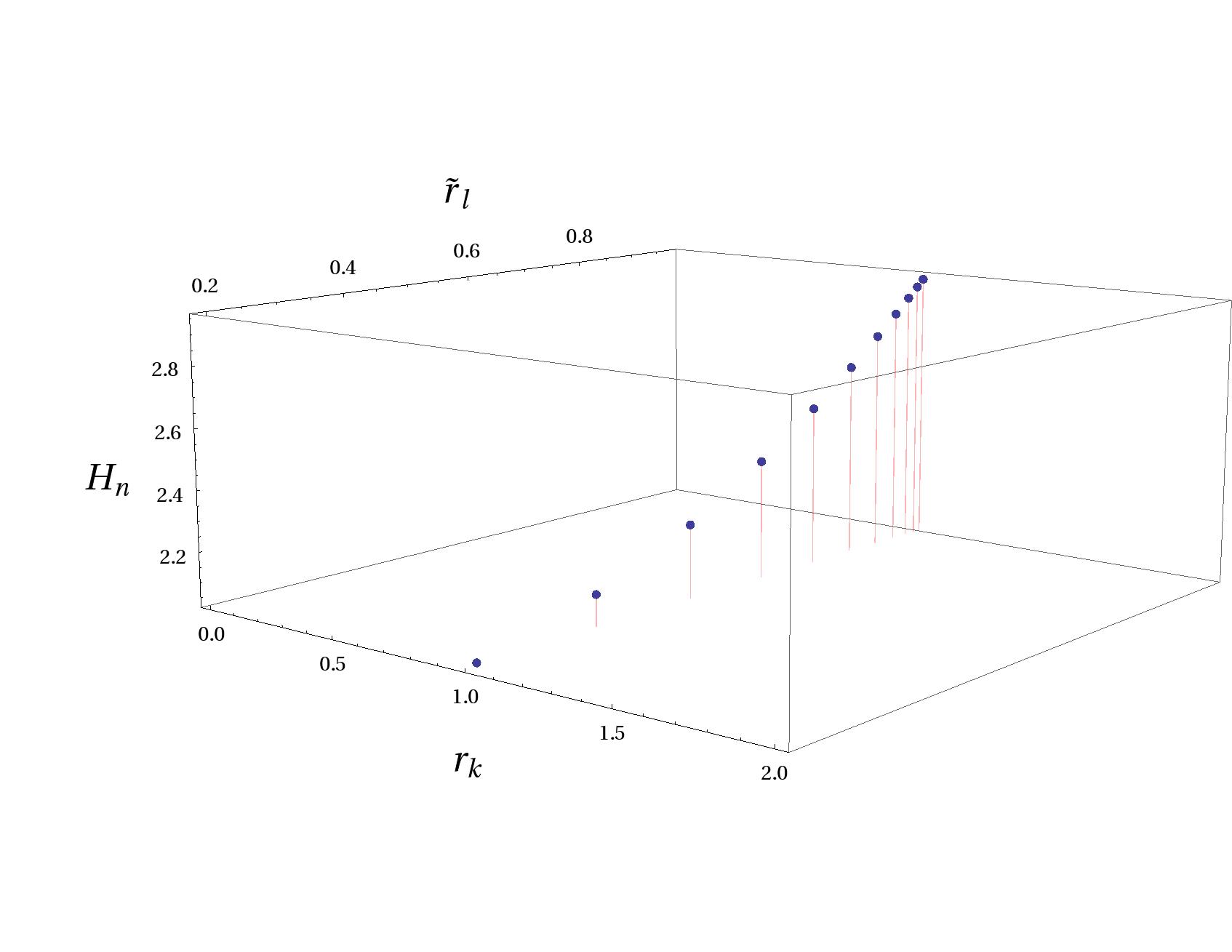}\caption{Graphic of the Hamiltonian function showing a discrete behaviour for the case  $\sqrt{\omega_k}r_k > \sqrt{\omega_l}\tilde{r}_l$ with $k=\omega_k=1$, $l=\omega_l=1$, $H_0=1$ and $r_k = 1$, showing a discrete behavior.}\label{tophamilfig3d1}
\end{figure}

It is interesting to notice that the spectrum of the Hamiltonian is not equidistant, an effect that can be interpreted as a result of the interaction of different modes 
of oscillation. For large values of $n$ the value of the Hamiltonian tends to a constant value 
$H_\infty = H_0 -  2\sqrt{\omega_k\omega_l}\, r_k \tilde{r}_l =H_0+2\omega_k r_k^2 $.
This behavior is illustrated in figure \ref{tophamilfigmesh} for both cases and in figure \ref{tophamilfig3d1} for $\sqrt{\omega_k}r_k > \sqrt{\omega_l}\tilde{r}_l$.

The complexity of the calculation rises as we increase the number of distinct modes of oscillation in a given direction. 
Nevertheless, based on the expressions given above for the spectra of different modes and the simple final form of the Hamiltonian, one can expect similar results for other more complicated configurations.

\section{Discussion}
\label{sec:disc}

We consistently developed the method to obtain the topological spectrum for a bosonic string moving in a general background. We computed the spectra for some particular configurations in the case of a Minkowski background space.
 The results are discrete relations between the amplitude of the modes of oscillation that describe the dynamics of the string. These relations account for an allowed set of embeddings of the worldsheet in the background spacetime. That is, the solutions $X^\mu=X^\mu(\tau,\sigma)$ which describe the embedding of the string in the spacetime are parametrized by a set of numbers constituting the modes of oscillation; then, in principle, one might be tempted to believe that any combination of these numbers may be realized. The topological spectrum sets restrictions to the values that these numbers can take, in particular the amplitudes of the modes of oscillation, thus limiting the cases of valid embeddings, which in our perspective seems very interesting. As a consequence, the Hamiltonian corresponding  to the energy of the worldsheet becomes a discrete quantity that corresponds to each allowed embedding. 
 
Due to the complexity of the computations, the above discretization was performed only for a limited number of oscillations. Nevertheless, the symmetry of the expressions for the solution and the topological spectrum allows us to conjecture the behavior of the discreteness in general. In fact, the general Hamiltonian for a closed string can be shown to be
\begin{equation}   
H = H_0 + \sum_I \sum_k \omega_k (r_k^I)^2 + \sum_J \sum_l \omega_l (\tilde{r}_l^J)^2 \ .
\label{gham}
\end{equation} 
Then,  we can infer the general spectrum
\begin{equation}   
\frac{4}{\pi}\, \ \prod_{kl} \omega_k\omega_l\, \ln \left[ \frac{\sum_I\left(\sum_k \sqrt{\omega_k}r_k^{I} + \sum_l\sqrt{\omega_l}\tilde{r}_l^{I}\right)^2 }
{\sum_I \left(\sum_k \sqrt{\omega_k}r_k^{I} - \sum_l\sqrt{\omega_l}\tilde{r}_l^{I}\right)^2 } \right] = n \ ,
\label{gts}
\end{equation} 
which reduces to the spectra (\ref{casonominspec1}), (\ref{ts2}), and (\ref{ts3}) in the corresponding limiting cases. Moreover, notice that if we consider the simple case of one single oscillation in only one direction, or one single oscillation in different directions, the expression inside the logarithm reduces to one, so that $n=0$ and no discretization appears. It then follows that oscillations in different transverse directions do not interact with each other. As soon as we consider a configuration with at least two different modes of oscillation in the same direction, the topological spectrum becomes nontrivial, leading to discrete relationships between different modes. 

The general spectrum (\ref{gts}) could be used to rewrite the general Hamiltonian (\ref{gham}) in such a way that the discreteness of the energy becomes plausible, as in the particular Hamiltonian (\ref{hamil1}). The final expression of the Hamiltonian, however, will depend on the relation between different modes of oscillation as, for example, given in Eq.(\ref{hamil2}).

One important result of the investigation of the topological spectrum of bosonic strings is that it does not depend on the background spacetime, in the sense that the expression for the spectrum depends only on the conformal factor of the induced metric which, in turn, can easily be derived, independently of the specific form of the background metric. This opens the possibility of investigating discretization conditions for bosonic strings moving on curved backgrounds in the same manner as described in the present work. This issue is currently under investigation. 
 
It would be interesting to compare the discretization conditions which follow from topological quantization with those that appear in the context of canonical quantization. However, this comparison is not yet possible. In fact, as mentioned before, two important elements of the quantization procedure are still lacking in the approach presented here, namely, the concepts of quantum states and quantum evolution. 

\section*{Acknowledgments}
This work was partially support by DGAPA-UNAM No. IN106110 and No. IN108309.
F. N. acknowledges support from DGAPA-UNAM (postdoctoral fellowship).


\appendix

\section{Details on the integration of the Euler form}
\label{app:uno}

\begin{figure}[htb]
\centering
\includegraphics[scale=0.18]{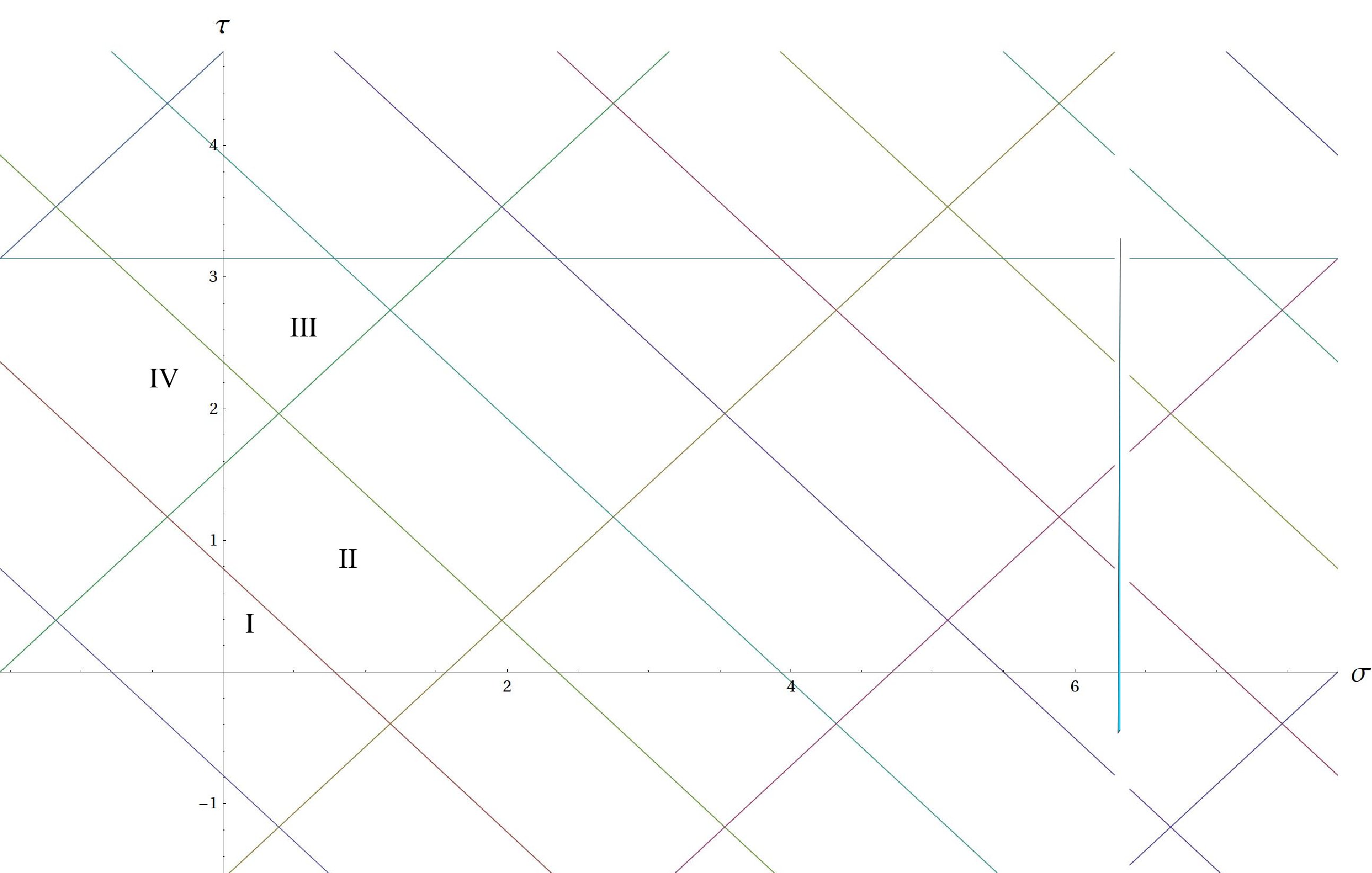}\caption{Domain of integration for the case of a right mode of oscillation with $k=\omega_k=1$ in the $J_1$ direction and left mode $l=\omega_l=2$ in the $J_2$ direction. Distinct regions are shown which correspond to the change of coordinates $I$ to $IV$.}\label{fig2}
\end{figure}

In order to integrate the Euler form (\ref{eulercasomin}) we must specify the limits in the domain of integration. For the parameter $\sigma$ the interval is $[0,2\pi]$ and is fixed, while for $\tau$ we notice that the expression for the Euler form is periodic in this parameter and we may choose a complete cycle. To perform the integral it is convenient to use null-like coordinates patches connected to the conformal coordinates by the following transformations,
\begin{align}  \label{coordcasomin}
\eta = x_I &= \sin \omega_k(\tau -\sigma +\gamma_k^{J_1}) \quad \text{and} \quad \xi = y_I = \sin \omega_l(\tau +\sigma +\tilde{\gamma}_l^{J_2}), \nonumber \\
x_{II} &= \sin \omega_k(\tau -\sigma +\gamma_k^{J_1}) \quad \text{and} \quad y_{II} = -\sin \omega_l(\tau +\sigma +\tilde{\gamma}_l^{J_2}), \nonumber \\
x_{III} &= -\sin \omega_k(\tau -\sigma +\gamma_k^{J_1}) \quad \text{and} \quad y_{III} = \sin \omega_l(\tau +\sigma +\tilde{\gamma}_l^{J_2}), \nonumber \\
x_{IV} &= -\sin \omega_k(\tau -\sigma +\gamma_k^{J_1}) \quad \text{and} \quad y_{IV} = -\sin \omega_l(\tau +\sigma +\tilde{\gamma}_l^{J_2}), 
\end{align}

\noindent where four types of regions are used to cover the whole domain of integration as seen in figure \ref{fig2}.

 The Euler form has the following aspect in this gauge,
\begin{equation} \label{eulercasominxy}
e(\mathcal{P}) = \pm \frac{2}{\pi}\frac{(r_k^{J_1}\tilde{r}_l^{J_2})^2\omega_k\omega_l x y}{\left[\omega_k\,(r_k^{J_1})^2 x^2 + \omega_l\,(\tilde{r}_l^{J_2})^2 y^2 \right]^2} dx \wedge dy,
\end{equation}

\noindent with the positive sign for regions $I$ and $IV$ and the negative one for $II$ and $III$. The parameters take values in the intervals $x \in [-1,1]$ and $y \in [-1,1]$. To cover the entire region of integration it is necessary to consider $2kl$ regions of the type $I$ and $IV$, and the same number of type $II$ and $III$.

In this case, for any type of region the integral of the Euler class vanishes,
\begin{equation}  \label{tscasominxy}
\int_{-1}^1 \int_{-1}^1 \, dxdy  \, e(x,y) = 0.
\end{equation}

This procedure to perform the integration is employed for the other particular configurations considered.

\bibliography{referencias}

\begin{thebibliography}{25}%
\makeatletter
\providecommand \@ifxundefined [1]{%
 \ifx #1\undefined \expandafter \@firstoftwo
 \else \expandafter \@secondoftwo
\fi
}%
\providecommand \@ifnum [1]{%
 \ifnum #1\expandafter \@firstoftwo
 \else \expandafter \@secondoftwo
\fi
}%
\providecommand \natexlab [1]{#1}%
\providecommand \enquote [1]{``#1''}%
\providecommand \bibnamefont  [1]{#1}%
\providecommand \bibfnamefont [1]{#1}%
\providecommand \citenamefont [1]{#1}%
\providecommand\href[0]{\@sanitize\@href}%
\providecommand\@href[1]{\endgroup\@@startlink{#1}\endgroup\@@href}%
\providecommand\@@href[1]{#1\@@endlink}%
\providecommand \@sanitize [0]{\begingroup\catcode`\&12\catcode`\#12\relax}%
\@ifxundefined \pdfoutput {\@firstoftwo}{%
 \@ifnum{\z@=\pdfoutput}{\@firstoftwo}{\@secondoftwo}%
}{%
 \providecommand\@@startlink[1]{\leavevmode\special{html:<a href="#1">}}%
 \providecommand\@@endlink[0]{\special{html:</a>}}%
}{%
 \providecommand\@@startlink[1]{%
  \leavevmode
  \pdfstartlink
   attr{/Border[0 0 1 ]/H/I/C[0 1 1]}%
   user{/Subtype/Link/A<</Type/Action/S/URI/URI(#1)>>}%
  \relax
 }%
 \providecommand\@@endlink[0]{\pdfendlink}%
}%
\providecommand \url  [0]{\begingroup\@sanitize \@url }%
\providecommand \@url [1]{\endgroup\@href {#1}{\urlprefix}}%
\providecommand \urlprefix [0]{URL }%
\providecommand \Eprint[0]{\href }%
\@ifxundefined \urlstyle {%
  \providecommand \doi [1]{doi:\discretionary{}{}{}#1}%
}{%
  \providecommand \doi [0]{doi:\discretionary{}{}{}\begingroup
  \urlstyle{rm}\Url }%
}%
\providecommand \doibase [0]{http://dx.doi.org/}%
\providecommand \Doi[1]{\href{\doibase#1}}%
\providecommand \selectlanguage [0]{\@gobble}%
\providecommand \bibinfo [0]{\@secondoftwo}%
\providecommand \bibfield [0]{\@secondoftwo}%
\providecommand \translation [1]{[#1]}%
\providecommand \BibitemOpen[0]{}%
\providecommand \bibitemStop [0]{}%
\providecommand \bibitemNoStop [0]{.\EOS\space}%
\providecommand \EOS [0]{\spacefactor3000\relax}%
\providecommand \BibitemShut [1]{\csname bibitem#1\endcsname}%
\bibitem[{\citenamefont{Alvarez}(1985)}]{alvarez}%
  \BibitemOpen
  \bibfield{author}{%
  \bibinfo {author} {\bibnamefont{Alvarez}, \bibfnamefont{O.}},\ }%
  \bibfield{title}{%
  \enquote{\bibinfo {title} {Topological quantization and cohomology},}\ }%
  \bibfield{journal}{%
  \bibinfo {journal} {Comm. Math. Phys.}\ }%
  \textbf{\bibinfo {volume} {100}},\ \bibinfo {pages} {279--309} (\bibinfo
  {year} {1985})\BibitemShut{NoStop}%
\bibitem[{\citenamefont{Bulgadaev}(2006)}]{bulgadaev}%
  \BibitemOpen
  \bibfield{author}{%
  \bibinfo {author} {\bibnamefont{Bulgadaev}, \bibfnamefont{S.~A.}},\ }%
  \bibfield{title}{%
  \enquote{\bibinfo {title} {Topological quantization of current in quantum
  tunnel contacts},}\ }%
  \bibfield{journal}{%
  \bibinfo {journal} {Pis'ma v Zh. {\`{E}}ksper. Teoret. Fiz.}\ }%
  \textbf{\bibinfo {volume} {83}},\ \bibinfo {pages} {659--663} (\bibinfo
  {year} {2006})\BibitemShut{NoStop}%
\bibitem[{\citenamefont{Carlip}(2001)}]{carlip}%
  \BibitemOpen
  \bibfield{author}{%
  \bibinfo {author} {\bibnamefont{Carlip}, \bibfnamefont{S.}},\ }%
  \bibfield{title}{%
  \enquote{\bibinfo {title} {Quantum gravity: a progress report},}\ }%
  \bibfield{journal}{%
  \bibinfo {journal} {Rep. Prog. Phys.}\ }%
  \textbf{\bibinfo {volume} {64}},\ \bibinfo {pages} {885--942} (\bibinfo
  {year} {2001})\BibitemShut{NoStop}%
\bibitem[{\citenamefont{Choi}(1994)}]{choi}%
  \BibitemOpen
  \bibfield{author}{%
  \bibinfo {author} {\bibnamefont{Choi}, \bibfnamefont{M.~Y.}},\ }%
  \bibfield{title}{%
  \enquote{\bibinfo {title} {Bloch oscillation and topological quantization},}\
  }%
  \bibfield{journal}{%
  \bibinfo {journal} {Phys. Rev. B}\ }%
  \textbf{\bibinfo {volume} {50}},\ \bibinfo {pages} {13875--13878} (\bibinfo
  {year} {1994})\BibitemShut{NoStop}%
\bibitem[{\citenamefont{Choquet-Bruhat}\
  \emph{et~al.}(1982)\citenamefont{Choquet-Bruhat},
  \citenamefont{DeWitt-Morette},\ and\ \citenamefont{Dillard-Bleick}}]{damas}%
  \BibitemOpen
  \bibfield{author}{%
  \bibinfo {author} {\bibnamefont{Choquet-Bruhat}, \bibfnamefont{Y.}}, \bibinfo
  {author} {\bibnamefont{DeWitt-Morette}, \bibfnamefont{C.}},\ and\ \bibinfo
  {author} {\bibnamefont{Dillard-Bleick}, \bibfnamefont{M.}},\ }%
  \emph{\bibinfo {title} {Analysis, Manifolds and Physics}}\ (\bibinfo
  {publisher} {Elsevier Science Publishers},\ \bibinfo {year}
  {1982})\BibitemShut{NoStop}%
\bibitem[{\citenamefont{Deguchi}(2007)}]{deguchi}%
  \BibitemOpen
  \bibfield{author}{%
  \bibinfo {author} {\bibnamefont{Deguchi}, \bibfnamefont{S.}},\ }%
  \bibfield{title}{%
  \enquote{\bibinfo {title} {Atiyah-{S}inger index theorem in an so(3)
  {Y}ang-{M}ills-{H}iggs system and derivation of a charge quantization
  condition},}\ }%
  \bibfield{journal}{%
  \bibinfo {journal} {Prog. Theor. Phys.}\ }%
  \textbf{\bibinfo {volume} {118}},\ \bibinfo {pages} {769--784} (\bibinfo
  {year} {2007})\BibitemShut{NoStop}%
\bibitem[{\citenamefont{Dirac}(1931)}]{dirac1931}%
  \BibitemOpen
  \bibfield{author}{%
  \bibinfo {author} {\bibnamefont{Dirac}, \bibfnamefont{P.~A.~M.}},\ }%
  \bibfield{title}{%
  \enquote{\bibinfo {title} {Quantised singularities in the electromagnetic
  field},}\ }%
  \bibfield{journal}{%
  \bibinfo {journal} {Proc. Roy. Soc.}\ }%
  \textbf{\bibinfo {volume} {A 133}},\ \bibinfo {pages} {60--72} (\bibinfo
  {year} {1931})\BibitemShut{NoStop}%
\bibitem[{\citenamefont{Frankel}(2004)}]{frankel}%
  \BibitemOpen
  \bibfield{author}{%
  \bibinfo {author} {\bibnamefont{Frankel}, \bibfnamefont{T.}},\ }%
  \emph{\bibinfo {title} {The Geometry of Physics}},\ \bibinfo {edition} {2nd}\
  ed.\ (\bibinfo {publisher} {Cambridge University Press},\ \bibinfo {year}
  {2004})\BibitemShut{NoStop}%
\bibitem[{\citenamefont{Isham}(1997)}]{isham2}%
  \BibitemOpen
  \bibfield{author}{%
  \bibinfo {author} {\bibnamefont{Isham}, \bibfnamefont{C.~J.}},\ }%
  \bibfield{title}{%
  \enquote{\bibinfo {title} {Structural issues in quantum gravity},}\ }%
  \bibfield{journal}{%
  \bibinfo {journal} {Gen. Rel. Grav.}\ }%
  \textbf{\bibinfo {volume} {GR14}},\ \bibinfo {pages} {167--209} (\bibinfo
  {year} {1997})\BibitemShut{NoStop}%
\bibitem[{\citenamefont{Johnson}(2003)}]{joh}%
  \BibitemOpen
  \bibfield{author}{%
  \bibinfo {author} {\bibnamefont{Johnson}, \bibfnamefont{C.~V.}},\ }%
  \emph{\bibinfo {title} {D-Branes}}\ (\bibinfo {publisher} {Cambridge
  University Press},\ \bibinfo {year} {2003})\BibitemShut{NoStop}%
\bibitem[{\citenamefont{Leone}\ and\ \citenamefont{L\'evy}(2008)}]{leone}%
  \BibitemOpen
  \bibfield{author}{%
  \bibinfo {author} {\bibnamefont{Leone}, \bibfnamefont{R.}}\ and\ \bibinfo
  {author} {\bibnamefont{L\'evy}, \bibfnamefont{L.}},\ }%
  \bibfield{title}{%
  \enquote{\bibinfo {title} {Topological quantization by controlled paths:
  Application to {C}ooper pairs pumps},}\ }%
  \bibfield{journal}{%
  \bibinfo {journal} {Phys. Rev. B}\ }%
  \textbf{\bibinfo {volume} {77}},\ \bibinfo {pages} {064524--064539} (\bibinfo
  {year} {2008})\BibitemShut{NoStop}%
\bibitem[{\citenamefont{Misner}(1978)}]{misner}%
  \BibitemOpen
  \bibfield{author}{%
  \bibinfo {author} {\bibnamefont{Misner}, \bibfnamefont{C.~W.}},\ }%
  \bibfield{title}{%
  \enquote{\bibinfo {title} {Harmonic maps as models for physical theories},}\
  }%
  \bibfield{journal}{%
  \bibinfo {journal} {Phys. Rev. D}\ }%
  \textbf{\bibinfo {volume} {18}},\ \bibinfo {pages} {4510--4524} (\bibinfo
  {year} {1978})\BibitemShut{NoStop}%
\bibitem[{\citenamefont{Naber}(1997)}]{naber}%
  \BibitemOpen
  \bibfield{author}{%
  \bibinfo {author} {\bibnamefont{Naber}, \bibfnamefont{G.~L.}},\ }%
  \emph{\bibinfo {title} {Topology, Geometry and Gauge Fields}}\ (\bibinfo
  {publisher} {Springer Verlag, New York},\ \bibinfo {year}
  {1997})\BibitemShut{NoStop}%
\bibitem[{\citenamefont{{}Nakahara}(2003)}]{nak}%
  \BibitemOpen
  \bibfield{author}{%
  \bibinfo {author} {\bibnamefont{{}Nakahara}, \bibfnamefont{M.}},\ }%
  \emph{\bibinfo {title} {Geometry, Topology and Physics}},\ \bibinfo {edition}
  {2nd}\ ed.\ (\bibinfo {publisher} {Taylor {\&} Francis},\ \bibinfo {year}
  {2003})\BibitemShut{NoStop}%
\bibitem[{\citenamefont{{}Nash}\ and\ \citenamefont{Sen}(1983)}]{nashsen}%
  \BibitemOpen
  \bibfield{author}{%
  \bibinfo {author} {\bibnamefont{{}Nash}, \bibfnamefont{C.}}\ and\ \bibinfo
  {author} {\bibnamefont{Sen}, \bibfnamefont{S.}},\ }%
  \emph{\bibinfo {title} {Topology and Geometry for Physicists}}\ (\bibinfo
  {publisher} {Academic Press},\ \bibinfo {year} {1983})\BibitemShut{NoStop}%
\bibitem[{\citenamefont{{}Nettel}\ and\
  \citenamefont{Quevedo}(2007)}]{netque2}%
  \BibitemOpen
  \bibfield{author}{%
  \bibinfo {author} {\bibnamefont{{}Nettel}, \bibfnamefont{F.}}\ and\ \bibinfo
  {author} {\bibnamefont{Quevedo}, \bibfnamefont{H.}},\ }%
  \bibfield{title}{%
  \enquote{\bibinfo {title} {Topological spectrum of classical
  configurations},}\ }%
  \bibfield{journal}{%
  \bibinfo {journal} {AIP Conf. Proc.}\ }%
  \textbf{\bibinfo {volume} {956}},\ \bibinfo {pages} {9--14} (\bibinfo {year}
  {2007})\BibitemShut{NoStop}%
\bibitem[{\citenamefont{{}Nettel}\ and\ \citenamefont{Quevedo}(2011)}]{netque}%
  \BibitemOpen
  \bibfield{author}{%
  \bibinfo {author} {\bibnamefont{{}Nettel}, \bibfnamefont{F.}}\ and\ \bibinfo
  {author} {\bibnamefont{Quevedo}, \bibfnamefont{H.}},\ }%
  \bibfield{title}{%
  \enquote{\bibinfo {title} {Topological quantization of the harmonic
  oscillator},}\ }%
  \bibfield{journal}{%
  \bibinfo {journal} {Int. J. of Pure and Appl. Math.}\ }%
  \textbf{\bibinfo {volume} {70}},\ \bibinfo {pages} {117--123} (\bibinfo
  {year} {2011})\BibitemShut{NoStop}%
\bibitem[{\citenamefont{{}Nettel}\ \emph{et~al.}(2009)\citenamefont{{}Nettel},
  \citenamefont{Quevedo},\ and\ \citenamefont{Rodr{\'i}guez}}]{netquemo}%
  \BibitemOpen
  \bibfield{author}{%
  \bibinfo {author} {\bibnamefont{{}Nettel}, \bibfnamefont{F.}}, \bibinfo
  {author} {\bibnamefont{Quevedo}, \bibfnamefont{H.}},\ and\ \bibinfo {author}
  {\bibnamefont{Rodr{\'i}guez}, \bibfnamefont{M.}},\ }%
  \bibfield{title}{%
  \enquote{\bibinfo {title} {Topological spectrum of mechanical systems},}\ }%
  \bibfield{journal}{%
  \bibinfo {journal} {Rep. Math. Phys.}\ }%
  \textbf{\bibinfo {volume} {64}},\ \bibinfo {pages} {355--365} (\bibinfo
  {year} {2009})\BibitemShut{NoStop}%
\bibitem[{\citenamefont{Pati{\~n}o}\ and\
  \citenamefont{Quevedo}(2003)}]{patquev2}%
  \BibitemOpen
  \bibfield{author}{%
  \bibinfo {author} {\bibnamefont{Pati{\~n}o}, \bibfnamefont{L.}}\ and\
  \bibinfo {author} {\bibnamefont{Quevedo}, \bibfnamefont{H.}},\ }%
  \bibfield{title}{%
  \enquote{\bibinfo {title} {Bosonic and fermionic behavior in gravitational
  configurations},}\ }%
  \bibfield{journal}{%
  \bibinfo {journal} {Mod. Phys. Lett. A}\ }%
  \textbf{\bibinfo {volume} {18}},\ \bibinfo {pages} {1331--1342} (\bibinfo
  {year} {2003})\BibitemShut{NoStop}%
\bibitem[{\citenamefont{Pati{\~n}o}\ and\
  \citenamefont{Quevedo}(2005)}]{patquev}%
  \BibitemOpen
  \bibfield{author}{%
  \bibinfo {author} {\bibnamefont{Pati{\~n}o}, \bibfnamefont{L.}}\ and\
  \bibinfo {author} {\bibnamefont{Quevedo}, \bibfnamefont{H.}},\ }%
  \bibfield{title}{%
  \enquote{\bibinfo {title} {Topological quantization of gravitational
  fields},}\ }%
  \bibfield{journal}{%
  \bibinfo {journal} {J. Math. Phys.}\ }%
  \textbf{\bibinfo {volume} {46}},\ \bibinfo {pages} {22502--22513} (\bibinfo
  {year} {2005})\BibitemShut{NoStop}%
\bibitem[{\citenamefont{Polchinski}(1998)}]{pol}%
  \BibitemOpen
  \bibfield{author}{%
  \bibinfo {author} {\bibnamefont{Polchinski}, \bibfnamefont{J.}},\ }%
  \emph{\bibinfo {title} {String Theory Vols. 1 and 2}}\ (\bibinfo {publisher}
  {Cambridge University Press},\ \bibinfo {year} {1998})\BibitemShut{NoStop}%
\bibitem[{\citenamefont{Ra{\~n}ada}\ and\
  \citenamefont{Trueba}(2006)}]{ranada}%
  \BibitemOpen
  \bibfield{author}{%
  \bibinfo {author} {\bibnamefont{Ra{\~n}ada}, \bibfnamefont{A.~F.}}\ and\
  \bibinfo {author} {\bibnamefont{Trueba}, \bibfnamefont{J.~L.}},\ }%
  \bibfield{title}{%
  \enquote{\bibinfo {title} {Topological quantization of the magnetic flux},}\
  }%
  \bibfield{journal}{%
  \bibinfo {journal} {Found. Phys.}\ }%
  \textbf{\bibinfo {volume} {36}},\ \bibinfo {pages} {427--436} (\bibinfo
  {year} {2006})\BibitemShut{NoStop}%
\bibitem[{\citenamefont{Schwarz}(1977)}]{schwarz}%
  \BibitemOpen
  \bibfield{author}{%
  \bibinfo {author} {\bibnamefont{Schwarz}, \bibfnamefont{A.~S.}},\ }%
  \bibfield{title}{%
  \enquote{\bibinfo {title} {On regular solutions of {E}uclidean {Y}ang-{M}ills
  equations},}\ }%
  \bibfield{journal}{%
  \bibinfo {journal} {Phys. Lett. B}\ }%
  \textbf{\bibinfo {volume} {67}},\ \bibinfo {pages} {172--174} (\bibinfo
  {year} {1977})\BibitemShut{NoStop}%
\bibitem[{\citenamefont{Zhong}\ and\ \citenamefont{Duan}(2008)}]{zhong}%
  \BibitemOpen
  \bibfield{author}{%
  \bibinfo {author} {\bibnamefont{Zhong}, \bibfnamefont{W.~J.}}\ and\ \bibinfo
  {author} {\bibnamefont{Duan}, \bibfnamefont{Y.~S.}},\ }%
  \bibfield{title}{%
  \enquote{\bibinfo {title} {Topological quantization of instantons in {SU}(2)
  {Y}ang-{M}ills theory},}\ }%
  \bibfield{journal}{%
  \bibinfo {journal} {Chin. Phys. Lett.}\ }%
  \textbf{\bibinfo {volume} {25}},\ \bibinfo {pages} {1534--1537} (\bibinfo
  {year} {2008})\BibitemShut{NoStop}%
\bibitem[{\citenamefont{Zwiebach}(2004)}]{zwiebach}%
  \BibitemOpen
  \bibfield{author}{%
  \bibinfo {author} {\bibnamefont{Zwiebach}, \bibfnamefont{B.}},\ }%
  \emph{\bibinfo {title} {A First Course in String Theory}}\ (\bibinfo
  {publisher} {Cambridge University Press},\ \bibinfo {year}
  {2004})\BibitemShut{NoStop}%
\end{thebibliography}%

\end{document}